\documentclass[12pt,preprint]{aastex}

\shortauthors{Ramsay Howat, Timmermann, Geballe, Bertoldi \& Mountain}
\shorttitle{Detection of rotational-vibrational emission from HD}

\begin{document}

\title{ Detection of a vibration-rotation emission line of
hydrogen deuteride toward Orion Peak 1: excitation coupling of HD
to H$_{2}$}

\author{S. K. Ramsay Howat} \affil{UK Astronomy Technology Centre, Royal
Observatory, Blackford Hill, Edinburgh, EH9 3HJ, UK.}
\email{skr@roe.ac.uk}

\author{R. Timmermann}

\affil{Universit\"at zu K\"oln, I. Physikalisches Institut, Z\"ulpicher
Str. 77, D-50937 K\"oln, Germany.}
\email{ralf.timmermann@epost.de}

\author{T. R. Geballe}

\affil{Gemini Observatory, 670 N. A'ohoku Place, Hilo HI 96720}

\author{F. Bertoldi}

\affil{Max-Planck Institut f\"ur Radioastronomie, Auf dem H\"ugel 69,
D-53121 Bonn, Germany.}

\and 

\author{C. M. Mountain}
\affil{Gemini Observatory, 670 N. A'ohoku Place, Hilo HI 96720}

\begin{abstract}

The 2.46~$\mu$m $v=1-0$ R(5) line of deuterated molecular hydrogen, HD,
has been detected in the Orion Peak~1 shock emission region, at a surface
brightness of $(8.5 \pm 2.1) \times 10^{-9}~\rm W m^{-2}sr^{-1}$ over a 6
arcsec$^2$ area. Comparison of the column density of HD($v=1,J=6$) with
the column density of HD($v=0,J=6$) previously observed from ISO and the H$_2$
level column densities toward the same region implies that the excitation
of HD is similar to that of H$_2$ for these energy levels, despite much higher spontaneous
transition rates for HD. We suggest that this rough equality is caused by
the coupling of the HD levels to those of H$_2$, due to strong reactive
collisions, $\rm HD + H \leftrightarrow H_2 + D$, in warm, partially
dissociated gas.  The deuterium abundance implied by the combined ISO and
UKIRT measurements toward Orion Peak 1 is [D]/[H]$=(5.1\pm 1.9)\times
10^{-6}$.

\end{abstract}


\section{Introduction}

Hydrogen deuteride (HD) is the simplest and most abundant deuterated
molecule. Its detection permits in principle a determination of the cosmic
deuterium abundance, and thereby places constraints on the physical
conditions during cosmic nucleosynthesis. Two pure rotational lines of HD
have been detected by the Short (SWS) and Long (LWS) Wavelength
Spectrometers on the Infrared Space Observatory (ISO). Wright et al.
(1999) detected the 0-0 R(0) transition (i.e., $v=0$, $J=1\rightarrow 0$)
line at 112~$\mu$m toward the Orion Bar photodissociation region, and from
an estimate of the H$_2$ column density and by assuming local
thermodynamic equilibrium (LTE) for HD, they determined a deuterium
abundance ratio, [D]/[H] $=(1.0\pm0.3) \times 10^{-5}$. Bertoldi et al.
(1999; hereafter B1999) reported a detection of the 0-0 R(5) line at
19.43~$\mu$m toward Orion Peak~1, the brightest position of shocked H$_2$
line emission in the Orion OMC-1 outflow (Beckwith et al. 1978).  They also derive upper limits
to the fluxes of sixteen pure rotational and vibration-rotation lines
toward the same position.  Combining their HD detection with observations
of H$_2$ lines by Rosenthal, Bertoldi \& Drapatz (2000; hereafter R2000),
they estimate [D]/[H]$~=~(7.6\pm2.9)\times10^{-6}$. Their estimate takes
into account the depletion of HD relative to H$_2$ that occurs in warm and
partially dissociated gas, from which much of the line emission is thought
to arise.

There are two major sources of uncertainty in the deuterium abundance
estimates derived from the mid-IR HD and H$_2$ emission lines.  One is the
possibility of variations of [HD]/[H$_2$] in warm, partially dissociated
gas, where the exothermic chemical reaction $\rm HD + H \rightarrow H_2 +
D$ depletes HD relative to H$_2$ (Timmermann 1996, 1998; B1999).  If the
dissociation fractions of HD and H$_2$ differ and vary along the line of
sight, the observations yield convolved averages over abundance and
excitation gradients, from which a deuterium abundance cannot be derived
directly.

The second major uncertainty is the derivation of an HD column density
from the measurement of but a single excited level.  In the absence of any
constraint on the HD excitation, B1999 initially assume that the HD giving
rise to the observed emission has the same excitation as the observed
post-shock H$_2$. Using this assumption, which essentially presumes that
both H$_2$ and HD are populated according to a LTE, they derive an HD
column density $N(\rm HD)_{\rm LTE} = (1.36\pm 0.38)\times10^{16}\rm
cm^{-2}$.  B1999 then note that, unlike H$_2$ levels at comparable energy,
HD(0,6) would not be expected to be populated according to LTE, mainly
because HD has a faster radiative deexcitation than H$_2$, and the
``critical density'' for HD(0,6) exceeds that which is expected to exist
in the emitting gas, $\sim 10^5 - 10^6\rm cm^{-3}$. From extensive modelling
of the HD level populations, B1999 derive a a factor of 1.5 (see section 3.4 of B1999)
to account approximately for the sub-LTE excitation of HD(0,6) and adjust
the total HD column upward.

This uncertainty in whether or not the HD level populations are in LTE,
and hence whether the column density should be corrected for non-LTE, can only
be resolved by measuring transitions from different HD levels. Here we
report the detection at Peak 1 of the HD 1-0 R(5) line at
2.46~$\mu$m. The upper level of this line is 7,747~K above ground
(calculated from the vibrational constants of Herzberg 1950 and the
rotational constants of Essenwanger \& Gush 1984), compared to 2,636~K
for the 0-0 R(5) line (B1999). Measurement of a second emission line of
HD at Peak~1 allows a first determination of the HD excitation in
shock-excited gas.

\section{Observations}

The observations were made at the United Kingdom Infrared Telescope
(UKIRT) on the night of 1999 January 20 UT, using the facility infrared
spectrometer, CGS4 (Mountain et al. 1990). The instrument contains a
256$\times$256 element InSb array. The echelle in CGS4 was used in
twenty-second order with a 2-pixel wide (0.82$\arcsec$) slit to provide a
spectral resolving power of $\sim$18,500 and wavelength coverage of
0.017~$\mu$m near 2.46~$\mu$m. The length of a pixel along the slit was
0.91$\arcsec$. The HD 1-0 R(5) line (2.458775~$\mu$m, Rich, Jones \&
McKellar 1982) is located at the long wavelength end of the K window where
the atmospheric transparency is generally poor; however, on Mauna Kea this
HD line is well isolated from nearby strong telluric lines.  The
atmospheric model spectrum generator, ATRAN (Lord 1992) was used to
determine this and it was confirmed by the observations.  The H$_2$ 1-0
Q(5) line at 2.45476~$\mu$m also falls in the observed wavelength range,
and is valuable for calibrating these observations with those done using
ISO.

The observations towards Peak~1 were made with the slit of CGS4 oriented
east-west and with row 84 of the array centered at $\alpha$(2000)$=$ 5$^h$
35$^m$ 13.7$^s$, $\delta$(2000)$=-$5$^{\circ}~$22\arcmin~8.5\arcsec. The
position was achieved by offsetting from near-by visible stars
and is accurate to better than 1arcsecond. This is the position for Peak
1 first identified by Beckwith et al (1978) and the ISO beam was centred
on this position (B1999). Following each on-source exposure of 120 seconds the telescope was offset
to a sky position 5 arcmin east of Peak~1. Sky and source frames were
repeated until the total on-source exposure time was 40 minutes.
Flat-field frames were also obtained. Individual exposures were
flat-fielded and sky-subtracted, and the sky-subtracted pairs were
despiked before being coadded. Wavelength calibration utilized telluric
lines in the observed spectrum of the standard star (wavelengths are
given in vacuo throughout this paper). The rms accuracy of
this calibration is estimated to be $5~{\rm km~s}^{-1}$.  The G4V star
HR~2007 ($K=4.44$, $T_{eff}=5740$~K assumed) was observed before and after
the observations of Peak~1; its spectrum was used to correct for telluric
features in the atmosphere and to provide flux calibration.  Weak stellar
features were removed from the spectrum of HR~2007 before ratioing, using
the solar spectrum (Livingston \& Wallace 1991) as a template.  No stellar
features are closely coincident with the HD line wavelength. The seeing
was poor during the observations and from the intensity profile of the
calibration star along the slit it is estimated that $50\pm10$\% of the
flux from the star passed through the CGS4 slit. This 20\% uncertainty is
included in the uncertainties in the line fluxes reported below. The
statistical uncertainties associated with the HD line are less than this.

\section{Results}

\subsection{Identification of the HD line}

The spectrum of Peak~1 is shown in Fig. 1. The H$_2$ 1-0 Q(5) line,
which dominates the spectrum, has a flux of
$(1.06\pm0.21)\times10^{-15}~{\rm W~m}^{-2}$, a full width at half
maximum (FWHM) of 0.00041~$\mu$m ($50\pm3~{\rm km~s}^{-1}$) and a full
width at 10 percent of 0.0018~$\mu$m ($220~{\rm km~s}^{-1}$). The broad
line widths are due to the fact that the H$_2$ in this region is
predominantly excited by shocks, as has long been established (e.g.,
Nadeau, Geballe, \& Neugebauer 1982). The resolution of the spectrum is
$16~{\rm km~s}^{-1}$, and thus the deconvolved widths are only
marginally less than the above values. The FWHM is in close agreement
with that previously observed for the H$_2$ 1-0 S(1) line (Nadeau \&
Geballe 1979; Chrysostomou et al. 1997; Stolovy et al. 1998).

The much fainter HD 1-0 R(5) line is clearly detected; its centroid is
$2.45922\pm0.00007 \mu$m (in vacuo). It is almost 1000 times fainter
than the H$_2$ Q-branch line. The flux in the line is
$(1.2\pm0.3)\times10^{-18} {\rm W~m}^{-2}$ where about half of the uncertainties are due to
the systematic uncertainty in the flux calibration. Even in this small
bright region, the surface brightness is a factor of 5 lower than the
limits measured by ISO for other 1-0 band HD lines (B1999). The FWHM of
the HD line, estimated from the locations of the line shoulders, is
$68\pm15~{\rm km~s}^{-1}$ (the uncertainty is 1~$\sigma$), in bare
agreement with the H$_2$ 1-0 Q(5) line width. No clear central peak is
seen, unlike for the H$_2$ line, but this could be due to noise
fluctuations.

To confirm the identification of the weak line as due to HD, we have
checked the wavelength in two ways. First, from the laboratory wavelength
of the HD line, the LSR velocity of Peak~1 ($+8\pm10 {~\rm km~s}^{-1}$,
Chrysostomou et al. 1997) and the Doppler shift (+16 km s$^{-1}$) due to
the earth's orbital motion on the date of the observation, the peak of the
HD line is calculated to occur at a laboratory vacuum wavelength of
$2.45911\pm0.00009$~$\mu$m. This value is in good agreement with the
measured line centroid, which we expect should be shifted only marginally
($\sim 0.0001~\mu$m) from the peak, as in the case of the H$_{2}$ line.  
Second, the shift of the HD line between the laboratory and observed
wavelengths, 0.00044~$\mu$m, is the same sign and nearly the same
magnitude as the shift of the H$_{2}$ line (0.00031~$\mu$m). The
difference of these shifts is less than the resolution of the spectrum.  
Moreover, while extensive line lists reveal a few atomic and ionic lines
(Ca {\sc I}, Cu {\sc III}, P{\sc iv}, Sc {\sc I}) with wavelengths within
0.0001~$\mu$m of the HD laboratory wavelength, none would be expected to
have a spatial distribution roughly mimicking that of H$_{2}$. To test
this, we examined the eight rows adjacent to the central eight brightest
rows, both to the east and west along the CGS4 slit. The ratio of the HD
line to the H$_2$ line is the same to within the estimated uncertainty of
30\%. These tests demonstrate that the newly detected line indeed is the
HD 1-0 R(5) transition.

\placetable{tab-1}

\subsection{Comparison with ISO: beam dilution}

The H$_2$ 1-0 Q(5) and HD 1-0 R(5) lines observed at UKIRT can be
compared with previous ISO observations centred on the same region
(R2000).  The CGS4 and ISO surface brightnesses and beam sizes are
summarised in Table 1.  The ground-based observation yields a surface
brightness which is (2.1$\pm$0.4) times larger than the
value measured with ISO.

The difference between the UKIRT and ISO measurements can be attributed to
the greatly different apertures of the two measurements. Whereas the ISO
observation averaged the emission over a large aperture centered on
Peak~1, the CGS4 flux was determined from the brightest eight adjacent
rows in the Peak 1 spectrum, a solid angle of just 6~arcsec$^2$. To
compare the observed flux from the HD 0-0 R(5) line observed by ISO and
the HD 1-0 R(5) line, we compensate for the different beam sizes using the
difference in the H$_2$ line fluxes as a beam dilution factor. In using
the difference in the H$_2$ fluxes to correct the HD line, we are assuming
that the H$_2$ and HD emissions have the same spatial distributions and
that the portions of Peak~1 sampled by CGS4 and by ISO have the same
excitation. The rough similarity the spatial distribution of HD and H$_2$
line emissions along the slit was already pointed out in Section~3.1.
Rosenthal et al. (2000) attribute only $\sim$5\% of the emission in the
ISO beam to the PDR bordering the foreground Orion Nebula H{\sc ii} region
and it is unlikely that the percentage has large variations within the ISO
beam. Therefore one can safely assume CGS4 slit samples H$_2$ with the
same excitation as the ISO beam.

The HD 0-0 R(5) flux measured with ISO was averaged over an even larger
aperture of 380 arcsec$^2$, so that there could be a modest difference in
beam dilution factors between the aperture in which the H$_2$ 1-0 Q(5)
line flux was derived and the HD 0-0 R(5). We ignore this possibility, and
for a comparison of the HD 0-0 R(5) and 1-0 R(5) line fluxes, we adopt a
beam dilution factor of 2.1, with a 20\% uncertainty. In other words, we
decrease the UKIRT HD line surface brightness by a factor of 2.1 for comparison with the
ISO HD line.

\subsection{HD column density and conditions in the post-shock gas}

The HD 1-0 R(5) average brightness observed in the 6 arcsec$^2$ aperture
can be converted to an average HD column density for the level from which
this transition arises, $v=1$, $J=6$, through 
\begin{equation} 
N(v,J)~=~{4\pi\over h c}~{\lambda~I_{\rm obs}\over A}~10^{0.4 A_\lambda},
\end{equation} 
where $A$ is the Einstein-$A$ coefficient (Table~\ref{tab-2}), and
$A_{\lambda}$ is the line of sight extinction at the line wavelength,
$\lambda$.  Adopting the values in Table~1 and an extinction of 0.78 mag
at 2.455~$\mu$m, which was derived by B1999 and R2000 (their Table~3) from
the relative H$_2$ line intensities, the observed line flux yields

\begin{equation}
N_{\rm obs}(1,6)  ~=~ (5.1 \pm 1.3) \times 10^{12}~ \rm cm^{-2}.
\end{equation}

To compare this with the HD $N(0,6)$ column density derived by B1999, we
correct the UKIRT column by $2.1\pm 0.4$ as justified above, obtaining
$N(1,6) = (2.4\pm 0.8)\times 10^{12}\rm cm^{-2}$, compared with $N(0,6) =
(3.0\pm 1.1)\times 10^{14}\rm cm^{-2}$ found by B1999.

Figure 2 is an amended version of Fig.~6 from B1999, now including our
new measurement. The figure includes a line which the HD level column
densities follow {\it if they have the same excitation as H$_2$}. There
is a good agreement of our new measurement with the prediction made by
this line, which indicates that the relative excitation of the HD(0,6)
and HD(1,6) levels is the same as that we find for H$_2$ levels at
similar energies. This is rather surprising, since one might have
expected that the population of these high HD levels would drop below
those of equivalent H$_2$ levels, due to the faster radiative
deexcitation rates of HD compared to H$_2$.  The radiative transition
probabilities of HD are much larger than those of H$_{2}$ because,
unlike H$_{2}$, ro-vibrationally excited HD can decay through the
emission of electromagnetic dipole radiation.  The total radiative decay
rate of the HD(1,6) level, e.g., is $5\times 10^{-5}\rm s^{-1}$, whereas
that of H$_{2}$(1,4), which is at comparable excitation energy, is about
60 times smaller. With comparable collisional excitation rates, the
level populations of H$_{2}$ and HD are expected to be quite different,
unless the gas density is very high, $n > 10^{7}$ cm$^{-3}$, which seems
unlikely.

\subsection{Coupling of the HD and H$_2$ excitations}

Why does the HD excitation shown by the levels (0,6) and
(1,6) appear to mimic the excitation of H$_2$ at similar level energies? 
We suggest that the cause may be excitation coupling between HD and
H$_2$. In shock-heated, partially dissociated gas the exchange reaction
\begin{equation}
  \label{eq:hdh2} \rm HD ~+~ H ~\leftrightarrow~ H_2 ~+~D~ 
\end{equation}
occurs. If the reaction time for this is faster than the time for
collisional and radiative equilibrium of HD energy levels to be
established, the HD level populations will be coupled to that of H$_2$
through the reactive collisions. Because the abundance of HD is much
smaller than that of H$_2$, the average H$_2$ level population would not
be significantly affected.

The HD radiative decay time is roughly $2\times 10^4$ seconds for the
$v=1$ levels. Vibrational relaxation of HD($v=1$)$\equiv \rm HD^*$ through
non-reactive collisions ($3 \times 10^{12} n^{-1}_{\rm H}$ sec and $3
\times 10^{11} n^{-1}_{\rm H}$ sec at 1000~K and 2000~K, respectively) is
slower than this at typical post-shock temperatures and densities. Thus it
is just the radiative time that must be compared with the reaction times.
There are two of the latter: the time that $\rm HD^*$ is
converted by the above forward reaction to H$_2$ and the time that HD$^*$
is formed from H$_2$ (through the reverse reaction).  If these times are
shorter, then the HD$^*$ abundance is driven by reactive collisions and
both molecules show nearly the same vibrational excitation, with HD
effectively becoming part of the H$_2$ level system.

The important reaction rates in eqn. (3) are those that
form HD$^*$. Because only a small fraction of the H$_2$ is vibrationally
excited, H$_2($v=0$)$+D $\rightarrow$ HD$^*$+H is the fastest production
channel of HD$^*$, even though per molecule H$_2$($v=1$) is more effective
(Timmermann 1996; Rozenshtein et al. 1985).  After employing the
results of Gray \& Balint-Kurti (1998) and Zhang \& Miller (1989), we
find a rate coefficient $k(2000\rm K) = 1.24\times 10^{-12}
~cm^{-3}s^{-1}$ for $\rm D+H_2(\it v=J={\rm
0})\rightarrow \rm HD^\ast+H$ at
2000~K. Per unit volume, the ``chemical excitation'' rate of
H$_2\rightarrow$HD$^*$ is then $k~n({\rm H_2}) n({\rm D})$, compared with
the HD$^*\rightarrow$HD radiative decay rate of $4\times10^{-5}~ n(\rm
HD^*) ~{\rm s}^{-1}$.  The formation of HD$^*$ is faster than
its decay when

\begin{equation}
n({\rm H_2}) ~>~ 4\times10^{7}~ {n({\rm HD})
\over n({\rm D})} {n({\rm HD^*})\over n({\rm HD})} ~{\rm cm}^{-3}.
\label{eq:condition}
\end{equation}

For gas
entering a partially dissociative C-type shock, the atomic hydrogen
abundance grows as the gas heats. The deuterium fraction [D]/[HD] can be 
computed from detailed balance of the HD-H$_2$ exchange reaction. Equation~(48) of Timmermann (1996) should read
\begin{equation}
r_{{\rm HD+H}} = r_{{\rm H_2+D}}~~ {Q({\rm H_2})~Q({\rm D})\over Q({\rm
HD})~Q({\rm H})}~~ e^{-\Delta E_{0}/{\rm k}T},
\label{eq:balance}
\end{equation}
where $Q = Q_{\rm trans}$ $Q_{\rm vibr}$ $Q_{\rm rot}$ is
the total partition function. The H$_2(v=0)$+D system lies $E_{0}/k$=418~K
above that of HD($v=0$)+H, and the H$_2$($v=1$)+D system lies 1183 K above
that of HD($v=1$)+H.  For the HD($v=0$)$\rightarrow$H$_2$($v=0$) exchange
reaction, detailed balance yields
\begin{equation}
{[{\rm D}] / [{\rm HD}]}~=~2.3~ ([{\rm H}] / [{\rm H_2}])~~e^{-418{\rm
K}/T}.
\end{equation}
Then eq. (4) becomes
\begin{equation}
n({\rm H}) ~>~ 2\times10^{7}~ {n({\rm HD^*})\over n({\rm HD)}} ~~
  e^{418{\rm K}/T}~{\rm cm}^{-3}.
\label{eq:nhcrit}
\end{equation}

In the portion of the post-shock region where the bulk of the observed
$v=1-0$ line emission occurs, the temperature is likely to be
$\sim$2000~K. There the fraction of H$_2$ in the $v=1$ state is $\sim$0.05
and that of HD would be similar if it were coupled to H$_2$. However, the
temperature profile of the shocked gas is not well known. We therefore
also estimate the excited fraction from the data of B1999 and R2000, who
observed the H$_2$ level populations over the line of sight of
shock-heated and cooling gas, which includes the entire range of
temperatures.  We obtain $N({\rm H_2}(v=1))/N(\rm H_2)\sim 0.001$.  The
relevant abundance fraction should lie between these two extremes. We
adopt $n({\rm HD^*})/n({\rm HD)}$ = 0.01 in the region where much of the
observed HD emission arises.  The atomic hydrogen density above which the
HD$^*$ abundance is dominated by H$_2$-HD reactive collisions is then
$2\times10^5 \rm~ cm^{-3}$. For comparison, at 2000~K the critical H
density for LTE excitation of HD$^*$ is $1.5\times 10^7 ~\rm cm^{-3}$.

In Orion the density of atomic hydrogen in C-shocks is believed to be well
in excess of $10^5~\rm cm^{-3}$ in the front (Timmermann 1996; Timmermann
1998; B1999). Thus partially dissociative shocks appear to be able to
couple HD strongly to H$_2$ over much of the warm shock layers.  In
regions with lower temperatures the exchange reactions are slower, but the
HD$^*$ fraction is also lower, so that the critical density
(eq.\ref{eq:nhcrit}) should not change much between 1000 K and 2000 K.

\section{Deuterium abundance}

We have found that the excitation of HD that is apparent through the
HD(0,6) and HD(1,6) levels is very similar to the excitation observed
(although averaged over a larger region) for H$_2$ at these level
energies.  Because of the lack of more information on the HD level
distribution, especially for the lower energy states $v=0$, $J=0-5$, we
assume that the HD and H$_2$ excitations are indeed similar.  Under
this assumption we can compute the HD partition sum to derive the total HD
column density.

Since HD is somewhat depleted relative to
H$_2$ in the warmest shock layers, the average excitation conditions of
HD might be somewhat different from those of H$_2$. The lower HD levels
are predominantly populated by cooler gas that does not couple to H$_2$,
but in this gas the low level populations may well follow LTE, so that
HD and H$_2$ would show a similar excitation.  Our assumption of
identical excitation for HD and H$_2$ is therefore a reasonable one.
Ideally, one would like to measure the level populations of the lower HD
levels, but the corresponding infrared lines are only accessible from
space with sufficient sensitivity. Upper limits derived from ISO
measurments (see Fig.~2) yield constraints only within a factor ten from
the adopted level populations.

Adopting the H$_2$ excitation measured by ISO, the total (warm) HD
column density is that given by B1999 as $N({\rm HD})_{\rm LTE} =
(1.36\pm 0.38) \times 10^{16} \rm cm^{-2}$, compared with $N({\rm H_2})
= (2.21\pm 0.24) \times 10^{21} \rm cm^{-2}$. Accounting for a 40\%
depletion of HD relative to H$_2$ in warm, partially dissociated gas
(see B1999 for a detailed justification of this factor), we  derive
a deuterium abundance [D]/[H]$=(5.1\pm 1.9)\times 10^{-6}$.  This value
is lower than the $(7.6\pm 2.9)\times 10^{-6}$ found by B1999, because
they expected a lower excitation of HD relative to that seen for H$_2$.

The deuterium abundance we derive is the lowest value yet measured.  Its
typical range derived through deuterium absorption measurments in the
local ISM is $(1-2) \times 10^{-5}$. Recent measurments by Jenkins et al.
(1999) and Sonneborn et al. (2000) however show that the abundance of
atomic deuterium does vary signifianctly along different lines of sight.
Our low value is comparable to the low value derived from absorption
measurments toward $\delta$ Ori A (Jenkins et al. 1999), which yielded
$7.4_{-1.3}^{+1.9}\times 10^{-6}$.

\section{Conclusion}

We have detected an emission line of vibrationally excited HD toward Orion
Peak 1.  The two HD lines now detected from this region arise from widely
different energy levels and their relative strengths surprisingly indicate
a similar excitation as H$_2$, in spite of the considerably different
spontaneous deexcitation rates of these isotopic molecular species. We
propose that the similarity in excitation is due to the strong
coupling of the HD and H$_2$ systems through reactive collisions and we
have revised the deuterium abundance at Peak~1 in view of the higher than
expected excitation of HD.

\acknowledgements

The United Kingdom Infrared Telescope is operated by the Joint
Astronomy Centre on behalf of the U. K. Particle Physics and Astronomy
Research Council.  We wish to thank the UKIRT Service Program and A.
Chrysostomou for obtaining these observations. We thank J.~Black for
initial discussions concerning the coupling of the HD-H$_2$
excitations. David Flower also contributed greatly during discussions of
this work. This paper makes use of the atomic line list data
maintained by Peter van Hoof and hosted by the Department of Physics and
Astronomy at the University of Kentucky
(www.pa.uky.edu/~peter/atomic). The authors are grateful to the referee
for comments on the originally submitted manuscript and for highlighting
recent work on HD reaction rates.

\clearpage

\begin{figure}
\epsscale{0.9}
\plotone{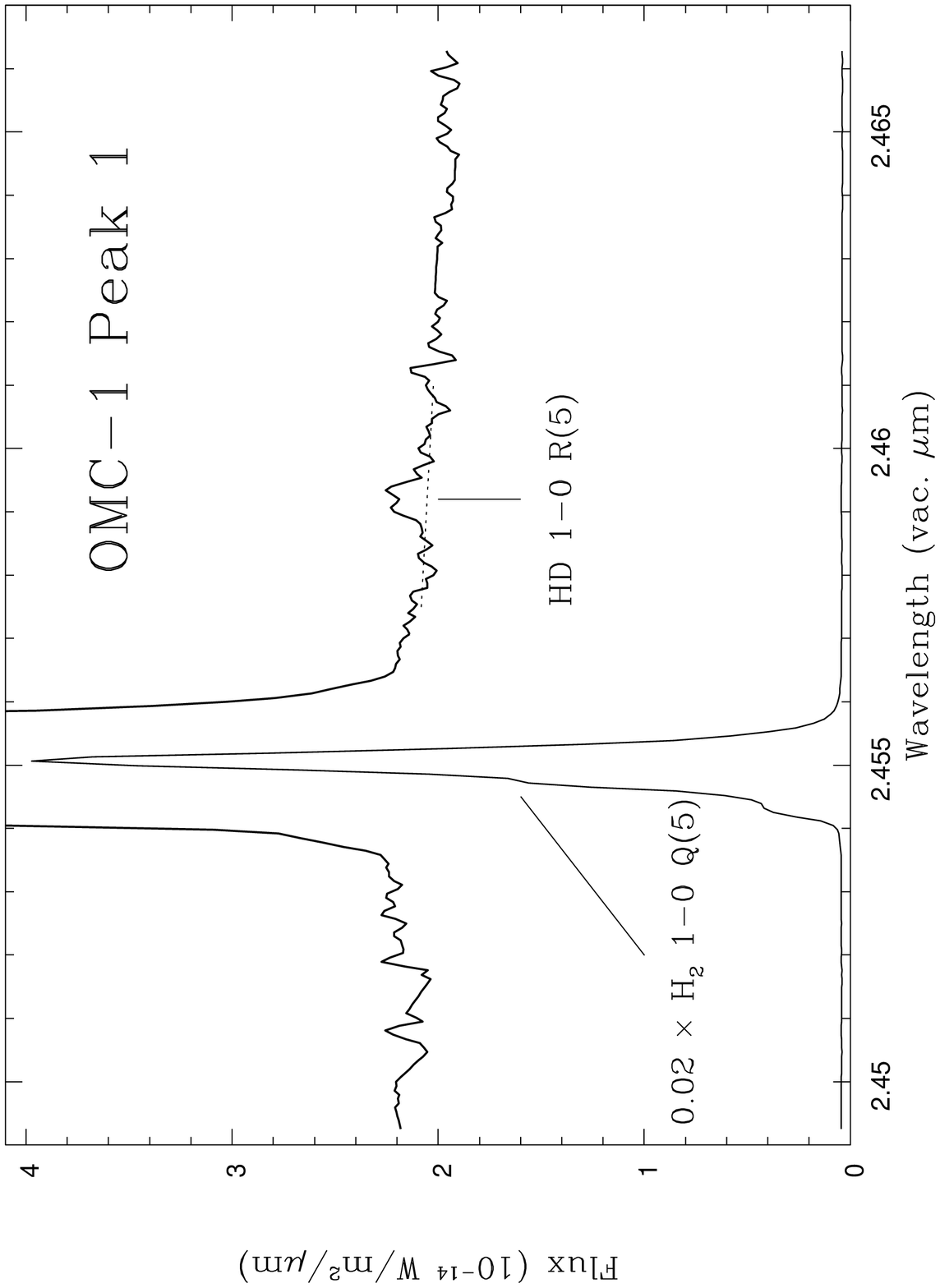}
\caption{Spectrum of a $0.82\arcsec\times 7.28\arcsec$ (NS
x EW) area of OMC-1 Peak~1 near 2.46~$\mu$m is shown in expanded form to reveal the
HD line, and also compressed by a factor of fifty to the peak intensity of
the H$_2$ 1-0 Q(5) line. The spectrum is a co-addition of the eight
brightest rows of H$_2$ line emission. The H$_2$ and HD lines are
indicated. The assumed continuum, used to calculate the flux
in the HD line, is also shown. The line flux was estimated by fitting
the continuum and integrating over a
0.0012~$\mu$m interval centered on the line. 
\label{fig1}. }
\end{figure}

\clearpage

\begin{figure}
\plotone{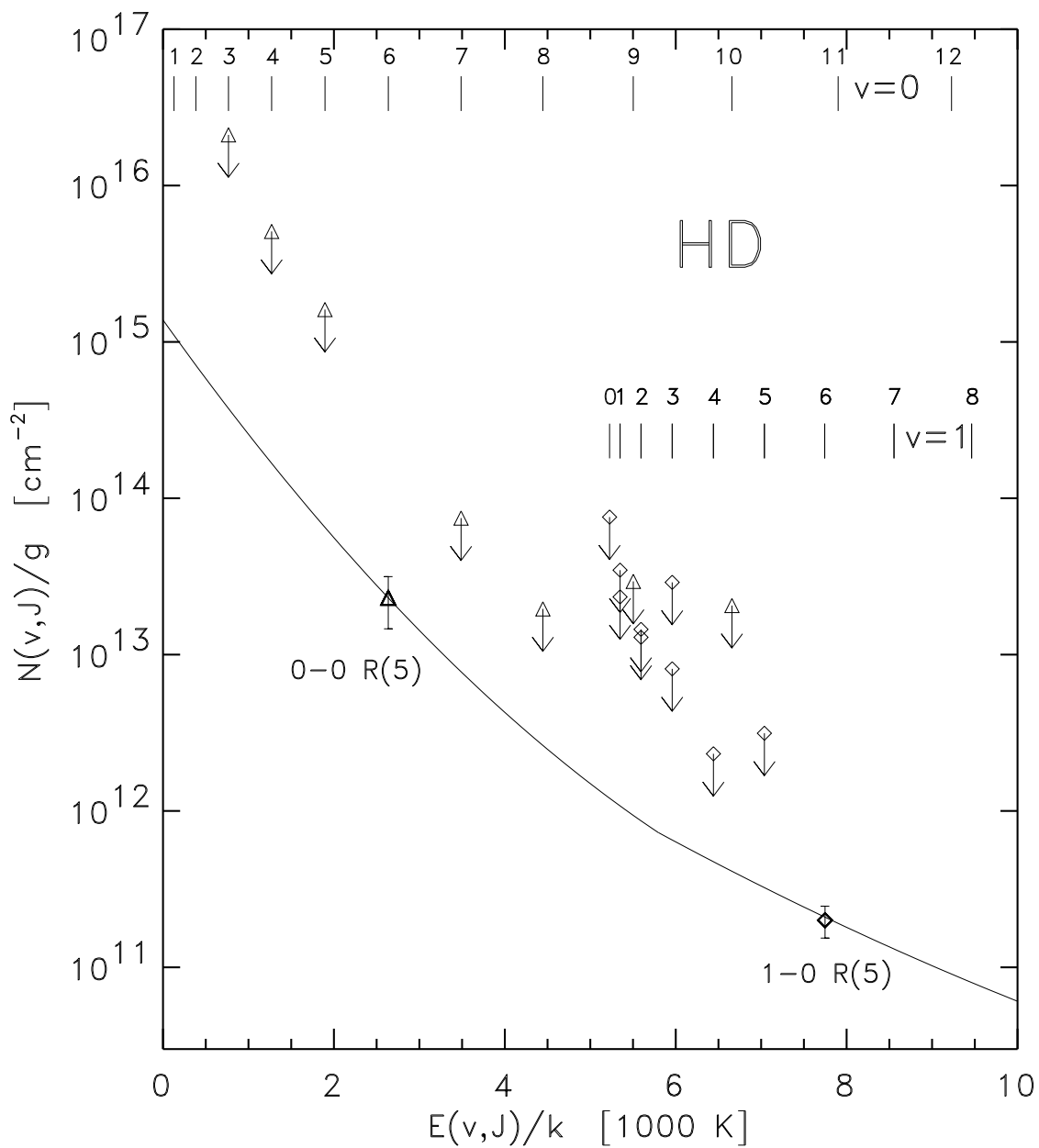}
\caption{Column densities of HD levels, divided by the level
degeneracy, plotted against the level energy. The $v=0$ column densities
or upper limits are indicated by triangles, the $v=1$ column densities by
squares. All but the HD(1,6) column density are from the B1999 ISO
measurements. The HD(1,6) column density is the value observed at UKIRT
divided by 2.1 for comparison with the ISO measurements, to correct for
the ISO beam dilution (see text). The solid line is a fit to the H$_2$
level populations (Rosenthal et al. 2000), normalised to the HD(0,6) column density.
 \label{fig2}}
\end{figure}

\noindent 
\clearpage

\begin{deluxetable}{lccccc}
\tablecaption{Line parameters for H$_{2}$ 1-0 Q(5) and HD
1-0 R(5) at OMC-1 Peak~1.\label{tab-1}}
\tabletypesize{\scriptsize}
\tablewidth{0pt}
\tablehead{\colhead{Line} & \colhead{$I_{\rm obs}$} & \colhead{beam} & \colhead{ $\lambda_{\rm obs}$ } &
\colhead{FWHM} & \colhead{$N(v,J)$ \tablenotemark{1}} \\
 & \colhead{$\rm W~m^{-2} sr^{-1}$} & \colhead{arcsec$^2$}&  \colhead{$\mu$m} & \colhead{$\mu$m} & \colhead{cm$^{-2}$}
}
\startdata
\\
\multicolumn{5}{c}{CGS4 Observations}\\
H$_2$ 1-0 Q(5) & $(7.6\pm 1.5)\times 10^{-6}$  & 5.97 &2.4550 &
$(4.1\pm0.2)\times10^{-4}$ & $(9.4 \pm
1.6) \times 10^{17}$ \\
 HD 1-0 R(5) & $(8.5 \pm 2.1) \times 10^{-9}$ & 5.97 & 2.4592 & $(5.6 \pm 1.2)
\times 10^{-4}$ & $ (5.1 \pm 1.3)
\times 10^{12} $ \\
\\
\hline
\\
\multicolumn{5}{c}{ISO-SWS Observations}\\
 H$_2$ 1-0 Q(5)\tablenotemark{2} & $(3.68 \pm 0.18) \times 10^{-6}$ & 280 & 2.4548 &1.6$\times 10^{-3}$ & $ (4.60 \pm 0.22) \times 10^{17} $ \\
 HD 0-0 R(5)\tablenotemark{3} & $(1.84 \pm 0.4) \times 10^{-8}$ & 380 &  19.4035 &   8.7$\times 10^{-3}$ & $ (3.1 \pm 1.1) \times 10^{14} $ \\

\tablenotetext{1} {Column density corrected for extinction (B1999)}
\tablenotetext{2} {From Rosenthal et al. (2000). The uncertainty in the
1-0 Q(5) line flux derives from
a 5\% flux calibration uncertainty (the line was detected with a signal
to noise ratio of 176). The line FHWM is approximate, calculated from the reported spectral resolving power of 1000-2000.}
\tablenotetext{3} {From Bertoldi et al. (1999)}
\enddata
\end{deluxetable}

\clearpage
\begin{deluxetable}{lccc}
\tablecaption{HD and H$_{2}$ transition parameters.\label{tab-2}}
\tablewidth{0pt}
\tablehead{
\colhead{Line} & \colhead{energy\tablenotemark{1}} &
\colhead{$A$ \tablenotemark{2}} &
\colhead{$g_J$ \tablenotemark{3}} \\
 & K & \colhead{sec$^{-1}$}
 & }
\startdata
\\
H$_2$ 1-0 Q(5) & 8365 & $2.55 \times 10^{-7}$ & 33  \\
 HD 1-0 R(5)  & 7747 & $ 5.29 \times 10^{-5}$  & 13 \\
 HD 0-0 R(5) & 2636 & $ 1.33 \times 10^{-5}$  & 13  \\

\tablenotetext{1}{HD energy level calculated from the vibrational
constants of Herzberg (1950) and the rotational constants of Essenwanger
\& Gush (1984)}
\tablenotetext{2} {HD transition probabilities from Abgrall, Roueff \&
Viala (1982)}
\tablenotetext{3} {Degeneracy of the upper level.
For the para-H$_2$ line the spin
degeneracy, g$_{\rm s}=3$, and g$_J$=g$_{\rm s}(2J+1)$}
\enddata
\end{deluxetable}


\begin{thebibliography}{}
\bibitem[Abgrall, H., Roueff, E. \& Viala, Y. 1982]{ab82} Abgrall, H.,
Roueff, E. \& Viala, Y. 1982, A\&AS, 50, 505

\bibitem[Bertoldi et al. 1999]{bert99} Bertoldi, F., Timmermann, R.,
Rosenthal, D., Drapatz, S. \& Wright, C.M. 1999, A\&A, 346, 267 (B1999)

\bibitem[Beckwith, 1978]{beck78} Beckwith, S., Persson, S.E., Neugebauer, G. and Becklin,
E.E. 1987, ApJ, 223, 464.

\bibitem[Brand et al. 1988]{bran88} Brand, P.W.J.L, Moorhouse, A., Burton,
M.G., Geballe, T.R., Bird, M. \& Wade, R. 1988, ApJL, 334, L103

\bibitem[Chrysostomou et al. 1997]{chrys97}Chrysostomou, A., Burton, M.G.,
Axon, D.J., Brand, P.W.J.L., Hough, J.H., Bland-Hawthorn, J. \& Geballe,
T.R. 1997, MNRAS, 289, 605

\bibitem[Essenwanger \& Gush 1984]{eg84} Essenwanger, P. \& Gush, H.P
1984, Can. J. Phys. 62, 1680

\bibitem[Flower \& Roueff 1999]{fr99} Flower, D.R. \& Roueff, E. 1999,
MNRAS, 309, 833

\bibitem[Gray \& Balint-Kurti 1998]{GB98} 
 Gray, S.K, Balint-Kurti, G.G. 1998, J.~Chem.~Phys. 108 (3), 950

\bibitem[Herzberg 1950]{h50} Herzberg, G. 1950, Molecular Physics and
Molecular Structure. Vol. I: Spectra of Diatomic Molecules, Van Nostrand
Reinhold, New York

\bibitem[Jenkins et al. 1999]{J99}
 Jenkins, E.B., Tripp, T.M., Wozniak, P.R., Sofia, U.J., Sonneborn, G., 
 1999, ApJ 520, 182 

\bibitem[Kress et al. 1990]{K90}
 Kress, J.D., Bacic, Z., Parker, G.A., Pack, R.T. 1990, J.~Phys.~Chem. 94,
 8055  

\bibitem[Lord 1992]{lord92} Lord, S.D., 1992, NASA Tech. Mem. 103957

\bibitem[Livingston \& Wallace 1991]{Liv91} Livingston, W. \& Wallace,
L. 1991, An Atlas of the Solar Spectrum from 1850 to 9000 cm$^{-1}$ (1.1
to 5.4 $\mu$m), N.S.O. Tech. Rep. \#91-101, National Solar Observatory

\bibitem[Mandy \& Martin 1993]{mm93} Mandy, M.E. \& Martin, P.G. 1993,
ApJS, 66, 199

\bibitem[Mountain et al. 1990]{mount90} Mountain, C.M., Robertson, D.J.,
Lee, T.J. \& Wade, R. 1990 SPIE Vol. 1235, D.L.Crawford (ed.), 25

\bibitem[Nadeau \& Geballe 1979]{nad79}Nadeau, D. \& Geballe, T.R. 1979,
ApJL 230, L169.

\bibitem[Nade, Geballe, \& Neugebauer 1982]{nad82}Nadeau, D., Geballe,
T.R., \& Neugebauer 1982, ApJ 253, 154

\bibitem[Rich, Jones \& McKellar1982]{rich82} Rich, Jones \& McKellar
1982, J. Mol. Spectr., 95, 432

\bibitem[Rosenthal et al. 2000]{rosen00} Rosenthal, D., Bertoldi, F. \&
Drapatz, S. 2000, A\&A, 356, 705 (R2000)

\bibitem[Rozenshtein et al. 1985]{roz85} Rozenshtein, V.B.,
Gershenzon, Yu, M., Ivanov, A.V., Il'in, S.D., Kucheryavii, S.I.,
\& Umanskii, S. Ya 1985, Chem. Phys. Lett., 121, 89

\bibitem[Sonneborn et al. 2000]{S00}
 Sonneborn, G., Tripp, T.M., Jenkins, E.B., Sofia, U.J., Vidal-Madjar,
 A., Wozniak, P.R. 2000, ApJ 545, 277 

\bibitem[Stolovy et al.]{stol98} Stolovy, S.R., Burton, M.G., Erickson,
E.F., Kaufman, M.J., Chrysostomou, A., Young, E.T., Colgan, S.W.J., Axon,
D.J., Thompson, R.I., Rieke, M.I. \& Schneider, G. 1998, ApJ, 492, L151

\bibitem[Timmermann 1996]{tim96} Timmermann, R. 1996, ApJ, 456, 631.

\bibitem[Timmermann 1998]{tim98} Timmermann, R. 1998, ApJ, 498, 246.

\bibitem[Ulivi et al. 91]{ulvi91} Ulivi, L., De Natale, P. \& Inguscio,
M. 1991, ApJL 378, L29

\bibitem[Wright et al. 1999]{wright99} Wright, C.M., van Dishoeck, E.F.,
Cox, P., Sidher, S., \& Kessler, M.F. 1999, ApJ, 515, L29.

\bibitem[Zhang \& Miller 1989]{ZM89} 
 Zhang, J.Z.H., Miller, W.H. 1989, J.~Chem.~Phys. 91 (3), 1528

\end{thebibliography}
\end{document}